\documentclass[prd,twocolumn,showpacs,superscriptaddress,nofootinbib,floatfix,showkeys,10pt]{revtex4-2}
\usepackage{bm}
\usepackage{times}
\usepackage{braket}
\usepackage{amsfonts,amssymb,stmaryrd,latexsym,amsmath}
\usepackage[usenames,dvipsnames]{color}
\usepackage{epsfig}
\usepackage{slashed}
\usepackage{hyperref}
\usepackage{subfigure}
\usepackage{graphics}
\usepackage{slashed}
\usepackage{orcidlink}
\usepackage{multirow}

\allowdisplaybreaks[1]

\definecolor{nicered}{rgb}{0.7,0.1,0.1}
\definecolor{nicegreen}{rgb}{0.1,0.5,0.1}
\definecolor{emph}{rgb}{1,0,0}
\definecolor{doub}{rgb}{0.7,0.2,1.0}
\definecolor{navyblue}{RGB}{0, 110, 184}

\hypersetup{colorlinks,citecolor=nicegreen,linkcolor=nicered,urlcolor=navyblue}

\begin{document}
	
\title{Unified description of the $Qs \bar q \bar q$ molecular bound states, molecular resonances and compact tetraquark states in the quark potential model} 

\author{Yan-Ke Chen\,\orcidlink{0000-0002-9984-163X}}\email{chenyanke@stu.pku.edu.cn}
\affiliation{School of Physics, Peking University, Beijing 100871, China}

\author{Wei-Lin Wu\,\orcidlink{0009-0009-3480-8810}}\email{wlwu@pku.edu.cn}
\affiliation{School of Physics, Peking University, Beijing 100871, China}

\author{Lu Meng\,\orcidlink{0000-0001-9791-7138}}\email{lu.meng@rub.de}
\affiliation{Institut f\"ur Theoretische Physik II, Ruhr-Universit\"at Bochum,  D-44780 Bochum, Germany }

\author{Shi-Lin Zhu\,\orcidlink{0000-0002-4055-6906}}\email{zhusl@pku.edu.cn}
\affiliation{School of Physics and Center of High Energy Physics,
Peking University, Beijing 100871, China}

\begin{abstract}

We calculate the mass spectrum of the $Qs\bar q \bar q$ $(Q=c, b)$ tetraquark states with $J^P=(0,1,2)^+$ using the AL1 quark potential model, which successfully describes the conventional hadron spectrum. We employ the Gaussian expansion method to solve the four-body Schr\"odinger equation, and use the complex scaling method to identify the resonances. With the notation $T_{Q s, I(J)}^{\text {Theo. }}(M)$,  we find several near-threshold bound states and resonances, including $T_{cs,0(0)}^{\mathrm{Theo.}}(2350)$, $T_{cs,0(0)}^{\mathrm{Theo.}}(2906)$, $T_{bs,0(0)}^{\mathrm{Theo.}}(5781)$, $T_{bs,0(1)}^{\mathrm{Theo.}}(5840)$, and $T_{bs,0(0)}^{\mathrm{Theo.}}(6240)$, which are close to the $D\bar{K}$, $D^*\bar{K}^*$, $\bar{B}\bar{K}$, $\bar{B}^*\bar{K}$, and $\bar{B}^*\bar{K}^*$ thresholds, respectively. Furthermore, their spatial structures clearly support their molecular natures. The resonance $T_{cs,0(0)}^{\mathrm{Theo.}}(2906)$ with a mass of $2906$ MeV, a width of $20$ MeV, and quantum numbers $I(J^P)=0(0^+)$ may serve as a good candidate for the experimental $T_{cs0}(2900)$ state. We strongly urge the experimental search of the predicted states.
\end{abstract}
 
\maketitle

\section{Introduction}~\label{sec:intro}
In 2020, the LHCb Collaboration observed two tetraquarks $T_{cs0,1}(2900)$ in the $D^{-} K^{+}$ invariant mass distribution in the process $B^{+} \rightarrow D^{+} D^{-} K^{+}$~\cite{LHCb:2020bls,LHCb:2020pxc},
\begin{equation}
\begin{aligned}
T_{cs0}(2900): & M=2866 \pm 7 \pm 2 \mathrm{MeV}, \\
& \Gamma=57 \pm 12 \pm 4 \mathrm{MeV}, \\
T_{cs1}(2900): & M=2904 \pm 5 \pm 1 \mathrm{MeV}, \\
& \Gamma=110 \pm 11 \pm 4 \mathrm{MeV}.
\end{aligned}
\end{equation}
The spin-parity quantum numbers are $J^P=0^+$ and $1^-$ for the $T_{c s 0}(2900)$ and $T_{c s 1}(2900)$, respectively. Their minimal quark contents are $cs\bar{u}\bar{d}$. These states might be the first tetraquark states with four different flavors observed in experiments, warranting further investigation into their existence and inner dynamics.

There are different ways to understand $T_{cs0,1}(2900)$. Molecular interpretations have been prompted by their masses close to the $D^*\bar K^*$ threshold~\cite{Molina:2010tx,Chen:2020aos,He:2020btl,Liu:2020nil,Hu:2020mxp,Agaev:2020nrc,Wang:2021lwy,Wang:2023hpp}. Alternative explanations such as the compact tetraquark states~\cite{Karliner:2020vsi,He:2020jna,Wang:2020xyc,Zhang:2020oze,Wang:2020prk,Lu:2020qmp,Tan:2020cpu,Albuquerque:2020ugi,Yang:2021izl,Agaev:2022eeh,Liu:2022hbk} and kinematic effects from the triangle singularities~\cite{Liu:2020orv,Burns:2020epm} have been proposed to elucidate the mass spectra and line shapes of $T_{cs0,1}(2900)$, respectively. Different aspects of the $T_{cs}$ states including their decays~\cite{Yu:2017pmn,Huang:2020ptc,Xiao:2020ltm}, productions~\cite{Chen:2020eyu,Lin:2022eau,Yu:2023avh}, and methods to find their partners~\cite{Bayar:2022wbx,Dai:2022qwh} have also been investigated. One can find more details in recent reviews~\cite{Brambilla:2019esw,Chen:2016qju,Guo:2017jvc,Liu:2019zoy,Meng:2022ozq,Chen:2022asf}.

The compact tetraquarks and hadronic molecular states have received the most attention among the explanations. It is crucial to perform dynamic calculations that treat the compact and molecular states equally, thereby discerning the actual configurations. Such computations can be accomplished by solving the four-body Schr\"odinger equation within the quark potential model~\cite{Meng:2021yjr,Meng:2023for}. In our previous work~\cite{Ma:2022vqf,Ma:2023int,Meng:2023jqk}, various quark models and few-body methods have been employed to perform the benchmark calculations for the tetraquark bound states, taking both dimeson and diquark-antidiquark spatial correlations into account. It was shown that the Gaussian expansion method is very efficient to investigate the tetraquark states. In this work, we choose a similar numerical method, and we also investigate the $Qs\bar{q}\bar{q}$ ($Q=c,b$) resonances.

We use the complex scaling method (CSM) to obtain the resonance solutions~\cite{Aguilar:1971ve,Balslev:1971vb}. In the CSM, the resonances and bound states can be treated similarly via the square-integrable basis expansion. It was well applied in atomic and molecular physics~\cite{PhysRevA.20.814,PhysRevA.55.4253,10.1063/1.464014}, nuclear physics~\cite{Myo:2001wn,Dote:2012bu,Papadimitriou:2014bfa,Maeda:2018xcl,Myo:2020rni}, and hadronic physics~\cite{Yang:2022bfu,Wang:2022yes} and was shown to be highly effective, see Refs~\cite{Moiseyev:1998gjp,aoyama2006complex,Carbonell:2013ywa} for reviews. Recently, the CSM in the momentum space has been adopted to study the exotic hadrons~\cite{Cheng:2022qcm, Lin:2022wmj, Lin:2023dbp}. An improved CSM to probe the coupled-channel virtual-state pole was introduced in Ref.~\cite{Chen:2023eri}.

Another innovation of this work is the definition of the root-mean-square (rms) radius between different quarks. Basically, the molecular or compact tetraquark states can be discerned through the analysis of their spatial configurations. The rms radius between different quarks is a commonly used criterion. However, the naive definition of the rms radius could be misleading when the antisymmetric wave function is required for the identical quarks. For instance, when the mesons $(Q\bar q)$ and $(s\bar q)$ form a molecular state, the wave function satisfying the Pauli principle is $|\psi_\mathcal{A}\rangle=|(Q\bar q_1)(s\bar q_2)\rangle-|(Q\bar q_2)(s\bar q_1)\rangle$. One can see that each light quark belongs to both mesons simultaneously. Therefore, neither $\langle\psi_\mathcal{A}|r^2_{Q\bar q}|\psi_\mathcal{A}\rangle$  nor $\langle\psi_\mathcal{A}|r^2_{s\bar q}|\psi_\mathcal{A}\rangle$ can reflect the size of the constituent mesons. In this study, we employ a new approach to define the rms radii, which can reflect the internal quark clustering behavior better, especially those of the molecular nature.


The paper is arranged as follows. The formulation is introduced in Sec.~\ref{sec:formulation}. In Sec.~\ref{sec:result}, the numerical results are discussed. At last, a summary is given in Sec.~\ref{sec:summary}.

\section{FORMULATION}\label{sec:formulation}

\subsection{Hamiltonian}\label{subsec:hamiltonian}

The nonrelativistic Hamiltonian for a four-quark system reads
\begin{equation}
  H=\sum_{j=1}^4 \frac{p_j^2}{2 m_j}-T_{\mathrm{c.m.}}+\sum_{i<j=1}^4 V_{i j}+\sum_j m_j,
\end{equation}
where the $m_i$ and $p_i$ are the mass and momentum of the quark $i$, respectively. The $T_{\mathrm{c.m.}}$ is the center-of-mass kinematic energy, which is subtracted to get the energies in the center-of-mass frame. The $V_{ij}$ represents the interaction between the (anti)quark pair $(ij)$. In this study, we adopt the AL1 quark potential model proposed in Refs.~\cite{Semay:1994ht,Silvestre-Brac:1996myf}:
\begin{equation}
  \begin{aligned}
V_{i j} =-\frac{3}{16} \boldsymbol\lambda_i^c \cdot \boldsymbol\lambda_j^c\left(-\frac{\kappa}{r_{i j}}+\lambda r_{i j}-\Lambda\right. \\
\left.+\frac{2 \pi \kappa^{\prime}}{3 m_i m_j} \frac{\exp \left(-r_{i j}^2 / r_0^2\right)}{\pi^{3 / 2} r_0^3} \boldsymbol{\sigma}_i \cdot \boldsymbol{\sigma}_j\right),
\end{aligned}
\end{equation}
where the $\boldsymbol\lambda^c_i$ and $\boldsymbol{\sigma}_i$ are the $\mathrm{SU}(3)$ color Gell-Mann matrix and $\mathrm{SU}(2)$ spin Pauli matrix acting on the quark $i$, respectively. In the conventional meson and baryon systems, the factor $\boldsymbol\lambda_i^c \cdot \boldsymbol\lambda_j^c$  is always negative and induces a confining interaction. However, in the tetraquark systems, the vanishing confinement terms  between color-singlet clusters allow the existence of the scattering states and possible resonances. The parameters of the model are taken from Ref.~\cite{Silvestre-Brac:1996myf}. These parameters were determined through a best-fit procedure applied to a comprehensive set of meson states across all flavor sectors. Therefore, the present work does not introduce any additional free parameters. The theoretical masses and rms radii of the corresponding mesons are listed in Table~\ref{tab:masses_radii_meson}. The theoretical results of the mesons are consistent with the experimental values up to tens of MeV, and we expect the errors for the tetraquark states to be of the same order.

\begin{table}[htbp]
    \centering
    \caption{The masses (in $\mathrm{MeV}$) and rms radii (in $\mathrm{fm}$) of the mesons in the quark model. The experimental results are taken from Ref.~\cite{ParticleDataGroup:2022pth}. ``Theo." and ``Expt." represent theoretical and experimental values, respectively.}
	 \label{tab:masses_radii_meson}
    \begin{tabular}{lccc|lcc}
    \hline\hline
        Mesons&$m_{\mathrm{Expt.}}$&$m_{\mathrm{Theo.}}$& $r^\mathrm{rms}_{\mathrm{Theo.}}$ & Mesons  &$m_{\mathrm{Theo.}}$&$r^\mathrm{rms}_{\mathrm{Theo.}}$ \\ \hline
        $K(1S)$ & 495.64 & 490.9  & 0.59 & $K(2S)$  & 1464.9  & 1.31 \\ 
        $K^*(1S)$ & 893.61 & 903.5  & 0.81 &  $K^*(2S)$   & 1642.7  & 1.42 \\ 
        $D(1S)$ & 1867.25 & 1862.4  & 0.61 & $D(2S)$   & 2643.2  & 1.23 \\
        $D^*(1S)$ & 2008.56 & 2016.1  & 0.70  &  $D^*(2S)$   & 2715.3  & 1.27 \\ 
        $B(1S)$ & 5279.50 & 5293.5  & 0.62   &$B(2S)$   & 6012.9  & 1.20 \\  
        $B^*(1S)$ & 5324.71 & 5350.5 & 0.66 & $B^*(2S)$    & 6040.6   & 1.21 \\ 
       \hline\hline
    \end{tabular}
\end{table}

\subsection{Calculation methods}\label{subsec:the_calculation_method}

\begin{figure}[htbp]
  \centering
  \includegraphics[width=0.9\linewidth]{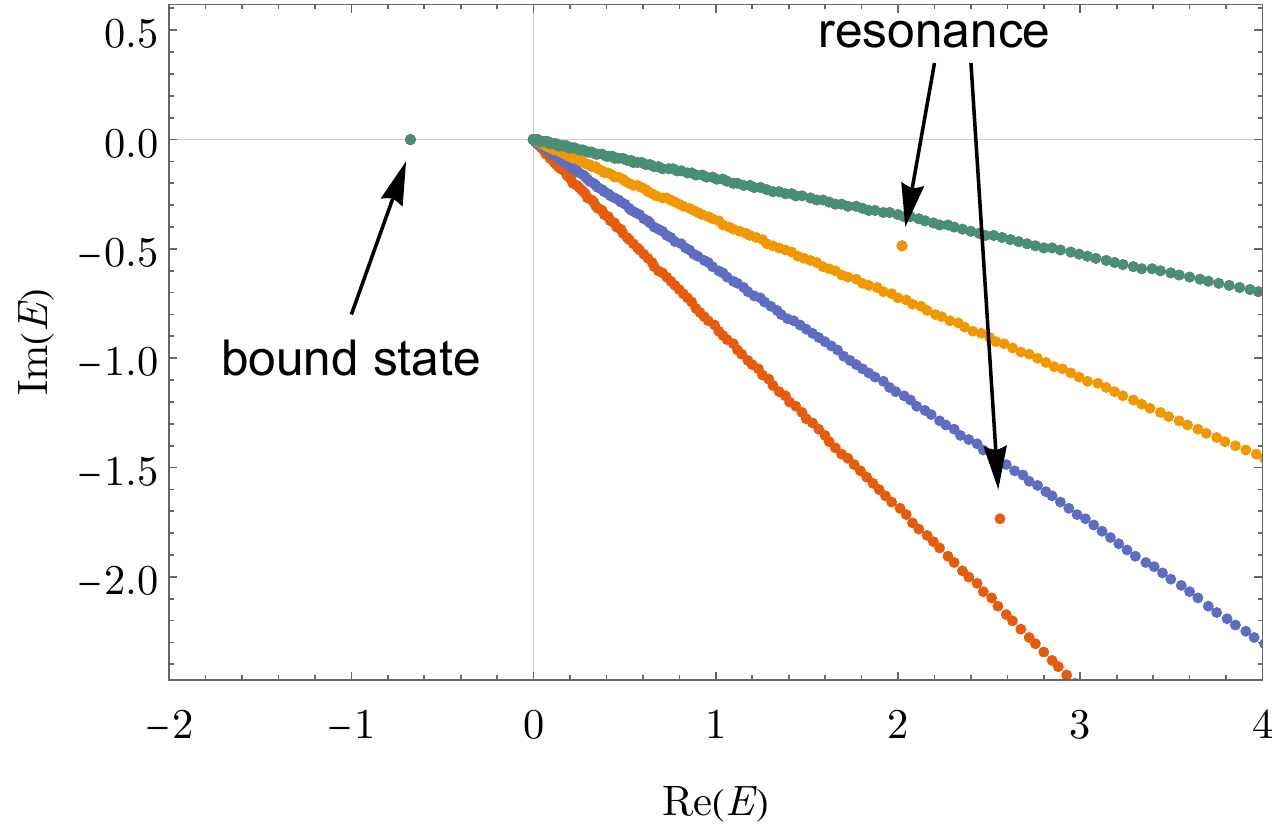}
  \caption{A typical solution of the complex-scaled Schr\"odinger equation.}
  \label{fig:csm_typical}
\end{figure}

In order to obtain the resonance, we apply the complex scaling method~\cite{Aguilar:1971ve,Balslev:1971vb}. The transformation $U(\theta)$ for the radial coordinate $r$ and its conjugate momentum $p$ is introduced as
\begin{equation}
  U(\theta) r=r e^{i \theta}, \quad U(\theta) p=p e^{-i \theta}.
\end{equation}
The Hamiltonian is transformed as
\begin{equation}
  H(\theta)=\sum_{j=1}^4 \frac{p_j^2e^{-2i\theta}}{2 m_j}+\sum_{i<j=1}^4 V_{i j}(r_{ij}e^{i\theta})+\sum_j m_j.
\end{equation}
Then the complex-scaled Schr\"odinger equation reads
\begin{equation}
  H(\theta)\Psi_{I,J}(\theta)=E(\theta)\Psi_{I,J}(\theta).
\end{equation}
A typical distribution of the eigenenergies solved by the CSM is shown in Fig.~\ref{fig:csm_typical}. The scattering states line up along $\operatorname{Arg}(E)=-2 \theta$. The bound states and resonances remain stable and do not shift as $\theta$ changes. The region enclosed by the continuum line and positive real axis correspond to the second Riemann sheet (RS-II), where the resonance pole can be detected from the CSM. The resonance with energy $E_R = m_R - i \frac{\Gamma_R}{2}$ can be detected when $2\theta>|\operatorname{Arg}(E_R)|$, where the $m_R$ and $\Gamma_R$ represent the mass and width of the resonance.

\begin{figure}[htbp]
  \centering
  \includegraphics[width=.9\linewidth]{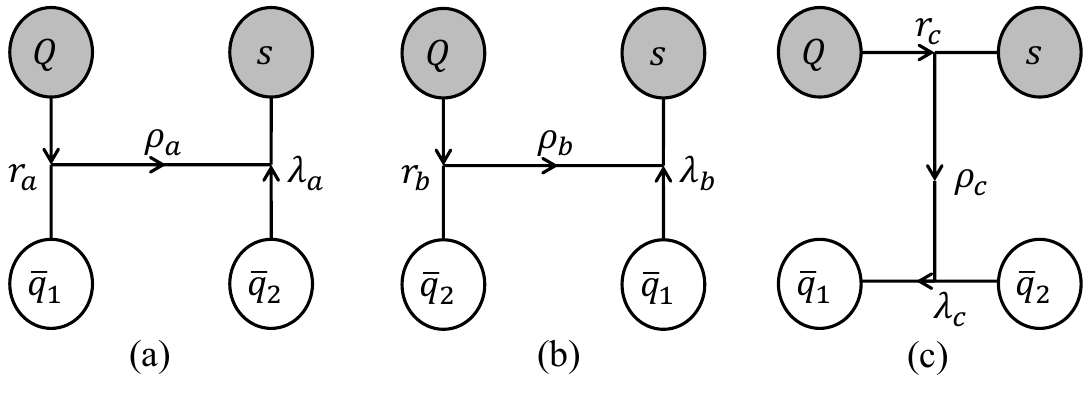}
  \caption{The Jacobi coordinates for two types of spatial configurations: (a),(b) for the dimeson configurations and (c) for the diquark-antidiquark configuration.}
  \label{fig:jac}
\end{figure}

The complex-scaled wave function of a $S$-wave $Qs\bar q \bar q$ tetraquark state can be expressed as
\begin{equation}\label{eq:wavefuntion}
\begin{aligned}
  \Psi_{I,J}(\theta)=\mathcal{A}\sum_{\alpha,n_\beta,\beta}&C_{\alpha,n_\beta,\beta}(\theta)\chi_\alpha^{I,J}\\
  &\times\phi^G_{n_{1,\beta}}(r_\beta)\phi^G_{n_{2,\beta}}(\lambda_\beta)\phi^G_{n_{3,\beta}}(\rho_\beta).
\end{aligned}
\end{equation}
where the $r_\beta, \lambda_\beta$, and $\rho_\beta$ are three independent Jacobi coordinates as shown in Fig.~\ref{fig:jac}. The superscript $\beta$ denotes the dimeson or the diquark-antidiquark spatial configurations. It is worth noting that one configuration is sufficient when all radially and orbitally excited basis functions are considered. However, a more efficient approach is to combine various $S$-wave configurations in the calculation~\cite{Hiyama:2003cu,Meng:2020knc,Meng:2021agn,Yang:2021izl,Deng:2022cld}. The $\mathcal{A}$ is the antisymmetric operator of the two light quarks. The $\chi_{\alpha}^{I,J}$ are the  spin-color-isospin wave functions with quantum numbers $(I,J)$, and they are given by
\begin{align}
\chi^{I,J}_{\bar 3_c,s_1,s_2}=\left[\left\{Qs\right\}_{\bar 3_c}^{0,s_1}~\left\{\bar q_1\bar q_2\right\}_{3_c}^{I,s_2}\right]_{1_c}^{I,J},\\
\chi^{I,J}_{6_c,s_1,s_2}=\left[\left\{Qs\right\}_{6_c}^{0,s_1}~\left\{\bar q_1\bar q_2\right\}_{\bar{6}_c}^{I,s_2}\right]_{1_c}^{I,J},\\
\chi^{I,J}_{1_c,s_1,s_2}=\left[\left\{Q\bar q_1\right\}_{1_c}^{\frac{1}{2},s_1}\left\{s\bar q_2\right\}_{1_c}^{\frac{1}{2},s_2}\right]_{1_c}^{I,J},\\
\chi^{I,J}_{8_c,s_1,s_2}=\left[\left\{Q\bar q_1\right\}_{8_c}^{\frac{1}{2},s_1}\left\{s\bar q_2\right\}_{8_c}^{\frac{1}{2},s_2}\right]_{1_c}^{I,J},
\end{align}
for all possible $|s_1,s_2,J\rangle$ compositions. The $C_{\alpha,n_\beta,\beta}(\theta)$ represent the undetermined expansion coefficients. We use the Gaussian expansion method to solve the four-body Schr\"odinger equation~\cite{Hiyama:2003cu}. The form of the spatial wave functions $\phi^G_{n_{1,\beta}}(r_\beta)$ is
\begin{equation}
    \begin{array}{c}
   \phi^G_{n_{1,\beta}}(r_\beta)=N_{n_{1,\beta}}e^{-\nu_{n_{1,\beta}}r_{\beta}^2},\\
   \nu_{n,\beta}=\nu_{1,\beta}\left(\frac{\nu_{n_{\mathrm{max}},\beta}}{\nu_{1,\beta}}\right)^{\frac{(n-1)}{(n_{\mathrm{max}}-1)}},
    \end{array}
\end{equation}
where $N_{n_{1,\beta}}$ is the normalization coefficients and $\nu_{n_{1,\beta}}$ is taken in a geometric progression. The spatial wave functions $\phi^G_{n_{2,\beta}}(\lambda_\beta)$ and $\phi^G_{n_{3,\beta}}(\rho_\beta)$ are similar.

\subsection{Spatial structures}\label{subsec:spatial_distribution}

Another important issue is to distinguish whether the exotic states are compact tetraquark states or hadronic molecules. In literature, the coefficients of the dimeson and diquark-antidiquark wave functions as depicted in Fig.~\ref{fig:jac} are often used as a criterion. It is essential to note that these two types of spatial configurations are not orthogonal. Therefore, the analysis and determination solely based on the proportion of a particular component in the spatial wave function are not straightforward. Another widely used criterion is the rms radius between different quarks. However, the conventional definition of the rms radius can be misleading because of antisymmetrization for the identical quarks in Eq.~\eqref{eq:wavefuntion}, as we point out in Sec.~\ref{sec:intro}. To avoid such ambiguities, we uniquely decompose the resulting antisymmetric wave function as follows:
\begin{equation}\label{eq:wf_decompose}
\begin{aligned}
  \Psi_{I,J}(\theta)=&|\{Q\bar q_1\}_{1_c}\{s\bar q_2\}_{1_c}\rangle\otimes|\psi_1^\theta)\\
  &+|\{Q\bar q_2\}_{1_c}\{s\bar q_1\}_{1_c}\rangle\otimes|\psi_2^\theta)
\end{aligned}
\end{equation}
In the conventional definition of the rms radius, the whole $\Psi_{I,J}(\theta)$ was used. Instead, we only use $|\psi_1^\theta\rangle$ to define the rms radius,
\begin{equation}\label{eq:rmsr}
  r^{\mathrm{rms}}_{ij}\equiv \mathrm{Re}\left[\sqrt{\frac{(\psi_1^\theta | r_{ij}^2 e^{2i\theta}|\psi_1^\theta)}{(\psi_1^\theta | \psi_1^\theta)}}\right].
\end{equation}
It is straightforward to verify that 
$$\mathcal{A}\left[\left|\left\{Q \bar{q}_1\right\}_{1_c}\left\{s \bar{q}_2\right\}_{1_c}\rangle \otimes | \psi_1^\theta\right)\right]=\Psi_{I,J}(\theta).$$
This definition helps to eliminate the potential confusion arising from the antisymmetrization. If the resulting $(Qs\bar{q}_1\bar{q}_2)$ state is a hadronic molecule, the $r^{\mathrm{rms}}_{Q\bar{q}_1}$ and $r^{\mathrm{rms}}_{s\bar q_2}$ are expected to be the sizes of the corresponding mesons, and should be smaller than the other radii in the four-body system. It is important to note that the inner products in the CSM are defined using the c product, accordingly~\cite{Romo:1968tcz},
\begin{equation}
  \left(\phi_n \mid \phi_m\right)\equiv\int r^2\phi_n(r)\phi_{m}(r)dr,
\end{equation}
where the square of the wave function rather than the square of its magnitude is used. The rms radius calculated by the c-product is generally not real. For the resonances that are not too broad, the real part of the rms radius can still reflect the internal quark clustering behavior, as discussed in Ref.~\cite{homma1997matrix}.

\section{Numerical Results}\label{sec:result}

\begin{figure*}[tbp]
  \centering
  \includegraphics[width=1\linewidth]{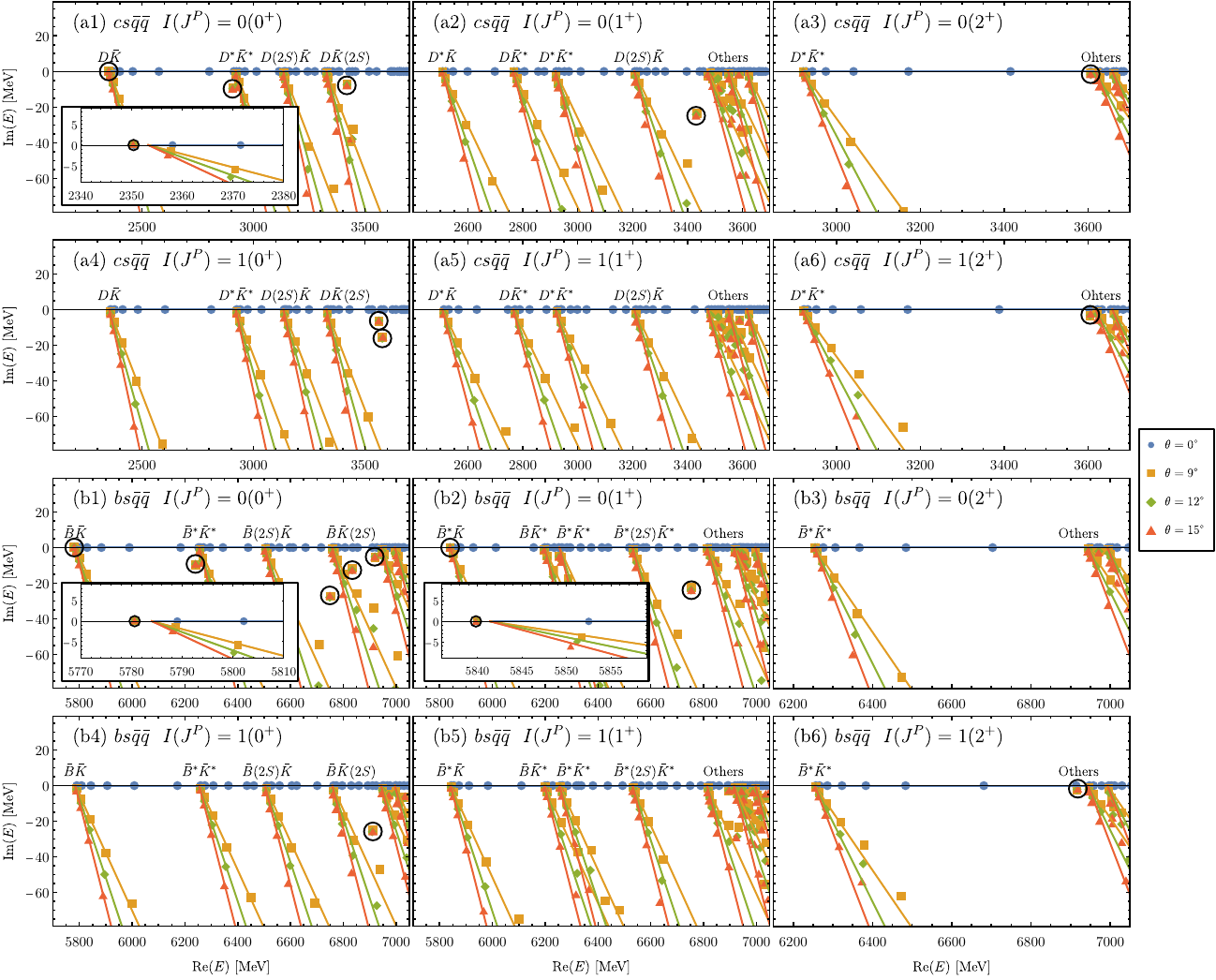}
  \caption{The complex energy eigenvalues of the (a) $cs\bar q\bar q$ states and (b) $bs\bar q\bar q$ states with varying $\theta$ in the CSM. The solid lines represent the continuum lines rotating along $\operatorname{Arg}(E)=-2 \theta$. The bound states and resonances remain stable and do not shift as $\theta$ changes.}
  \label{fig:Qsmass}
\end{figure*}

With the CSM, the complex eigenenergies of the $(cs\bar q\bar q)$ and $(bs\bar q\bar q)$ systems with various spin-parity quantum numbers are shown in Fig.~\ref{fig:Qsmass}. Most of the states rotate along the continuum lines, which correspond to the scattering states. The origin of the continuum lines on the real axis are the thresholds of the scattering states. The region enclosed by the continuum lines and positive real axis corresponds to the RS-II of the corresponding channels. The points that do not shift with $\theta$ represent the bound states and the resonances. Their eigenenergies and rms radii are summarized in Tables~\ref{tab:csmass_and_radii} and \ref{tab:bsmass_and_radii}, where the rms radii in the tables are calculated using Eqs.~\eqref{eq:wf_decompose} and \eqref{eq:rmsr}. For the resonances with $\Gamma>40$ MeV, their relatively large imaginary energy leads to numerically less accurate rms radii, and these radii should be considered as qualitative estimates. For convenience, we label the states in Tables~\ref{tab:csmass_and_radii} and \ref{tab:bsmass_and_radii} as $T_{Qs,I(J)}^{\mathrm{Theo.}}(M)$ in the subsequent discussion. 

It is important to reiterate that the present calculations may have an error at the order of tens of MeV originating from the quark model itself. Additionally, our calculations do not account for the widths of the constituent mesons, and only consider the two-body decays. Therefore, the theoretical widths of the resonances are expected to be smaller than the experimental values. Because of the large expected widths of the radially excited mesons, the results for the bound states and resonances below the $D^{*}\bar{K}^{*}(\bar{B}^{*}\bar{K}^{*})$ thresholds are more reasonable in the current calculations, which include $T_{cs,0(0)}^{\mathrm{Theo.}}(2350)$, $T_{cs,0(0)}^{\mathrm{Theo.}}(2906)$, $T_{bs,0(0)}^{\mathrm{Theo.}}(5781)$, $T_{bs,0(1)}^{\mathrm{Theo.}}(5840)$, and $T_{bs,0(0)}^{\mathrm{Theo.}}(6240)$. These states deserve further experimental investigation.

\begin{table}[htbp]
\centering
 \caption{The complex energies $E= M - i \Gamma/2$ (in MeV) and rms radii (in fm) of the $(cs\bar{q}\bar{q})$ states. The first four rows correspond to the ground-state meson-meson thresholds and the rms radii of the ground-state mesons, as taken from Table~\ref{tab:masses_radii_meson}. The $\Delta M$ represents the mass difference between the molecular states and their constituent mesons. The ``$\star$" labels the resonances with $\Gamma>40$ MeV, whose radii are  numerically less accurate and should be considered as qualitative estimates.}
 \label{tab:csmass_and_radii}
  \begin{tabular}{ccccccccc}
   \hline\hline
     States &$E$&$\Delta M$&$ r^{\mathrm{rms}}_{c\bar{q}_1} $  & $ r^{\mathrm{rms}}_{s\bar{q}_2} $  & $
   r^{\mathrm{rms}}_{cs} $ & $ r^{\mathrm{rms}}_{\bar{q}_1\bar{q}_2} $ & $ r^{\mathrm{rms}}_{c\bar{q}_2} $& $ r^{\mathrm{rms}}_{s\bar{q}_1} $\\\hline
    $D\bar K$ & $2353$ & & $0.61$ & $0.59$ \\
   $D^*\bar K$ & $2507$ & & $0.70$ & $0.59$\\
    $D\bar K^*$ & $2766$ & & $0.61$ & $0.81$\\
   $D^*\bar K^*$ & $2920$ & & $0.70$ & $0.81$\\
   \hline
    $ 0(0^+) $
   &$2350$ & $-3$ &  $ 0.61 $&$ 0.59 $&$ 2.45$&$ 2.52$&$ 2.47$&$ 2.50$\\
    &$ 2906-10i$& $-14$ &$ 0.74 $&$ 0.86 $&$ 1.12$&$ 1.26$&$ 1.21$&$ 1.27$\\
    & $ 3419-7i$ & & $0.91$& $1.09$ & $0.87$ & $1.22$ & $1.06$ & $1.09$ \\
    $ 0(1^+) $ & $ 3431-24i^{\star} $& & $1.11$&$ 1.26$&$ 0.57$&$ 0.79$&$ 1.10 $&$ 1.14 $\\
   $ 0(2^+) $ &$ 3607-2i $& & $ 0.83$&$ 1.30$&$ 1.10$&$ 1.50 $&$ 1.30 $&$ 1.26 $\\
   \hline
  $ 1(0^+) $  &$ 3563-6i $& & $ 0.89 $&$ 1.15 $&$ 0.97 $&$ 1.33 $&$ 1.10 $&$ 1.15$\\
    &$ 3578-16i $&&$ 0.99 $&$ 1.11 $&$ 0.95 $&$ 1.24 $&$ 1.03 $&$ 1.29 $\\
    $ 1(2^+) $  & $ 3605-3i $&&  $ 1.23 $&$ 1.17 $&$ 0.99 $&$ 1.29 $&$ 1.37 $&$ 1.38$\\
   \hline\hline
\end{tabular}
\end{table}

\begin{table}[htbp]
\centering
 \caption{The complex energies $E= M - i \Gamma/2$ (in MeV) and rms radii (in fm) of the $(bs\bar{q}\bar{q})$ states. The notations are the same as those in Table~\ref{tab:csmass_and_radii}.}
 \label{tab:bsmass_and_radii}
 \begin{tabular}{ccccccccc}
  \hline\hline
   States  &$E$ & $\Delta M$ &$r^{\mathrm{rms}}_{b\bar{q}_1}$ & $r^{\mathrm{rms}}_{s\bar{q}_2} $  & $
  r^{\mathrm{rms}}_{bs} $ & $ r^{\mathrm{rms}}_{\bar{q}_1\bar{q}_2} $ & $ r^{\mathrm{rms}}_{b\bar{q}_2} $& $ r^{\mathrm{rms}}_{s\bar{q}_1} $\\
  \hline
    $\bar{B}\bar K$ & $5784$ & & $0.62$ & $0.59$ \\
   $\bar{B}^*\bar K$ & $5841$ & & $0.66$ & $0.59$\\
   $\bar{B}\bar K^*$ & $6197$ & & $0.62$ & $0.81$\\
   $\bar{B}^*\bar K^*$ & $6254$ & & $0.66$ & $0.81$\\
  \hline
  $ 0(0^+) $ 
  &$5781$& $-3$ & $ 0.62 $&$0.59$&$ 2.11$&$ 2.21$&$ 2.14 $&$ 2.19 $\\
  &$ 6240-9i $ & $-14$ & $ 0.69 $&$ 0.86 $&$ 1.04 $&$ 1.21 $&$ 1.13 $&$ 1.21$\\
  &$6748-28i^{\star}$& & $1.12$ & $1.19$ & $0.74$ & $1.09$ & $1.14$ & $1.30$\\
 & $ 6834-13i $& & $ 0.77 $&$ 1.22$&$ 1.06 $&$ 1.23 $&$ 1.10 $&$ 1.24 $\\
  & $ 6920-5i $& & $ 0.72 $&$ 1.33 $ & $ 0.98 $ &$ 1.33 $ &$ 1.13 $ &$ 1.19 $\\
  $ 0(1^+) $  
  & $ 5840$&  $-1$ & $0.66 $& $0.59$ &$ 2.66 $&$ 2.75 $&$ 2.68 $&$ 2.73$\\
   & $ 6753-24i^{\star}$& &$ 1.11 $&  $ 1.18 $ &$ 0.82 $&$ 1.12 $&$ 1.21 $&$ 1.35 $\\
  \hline
  $ 1(0^+) $ 
  &$ 6909-26i^{\star} $ & & $ 1.04 $&$ 0.96 $&$ 0.72 $&$ 1.18 $&$ 0.86 $&$ 1.16$\\
  $ 1(2^+) $& 
  $ 6916-2i $ & &$ 1.18  $&$ 1.17 $&$ 0.79 $&$ 1.16 $&$ 1.23$&$ 1.28$\\
  \hline\hline
 \end{tabular}
\end{table}

\subsection{The $(cs\bar{q}\bar{q})$ sector}
For the $(cs\bar q\bar q)$ systems, we obtain a shallow bound state $T_{cs,0(0)}^{\mathrm{Theo.}}(2350)$ with a binding energy of $3$ MeV. As listed in Table~\ref{tab:csmass_and_radii}, the $r^{\mathrm{rms}}_{c\bar q_1}$ and $r^{\mathrm{rms}}_{s \bar q_2}$ of $T_{cs,0(0)}^{\mathrm{Theo.}}(2350)$ are very close to the radii of the $D$ and $\bar K$ mesons, respectively, which are significantly smaller than $(r^{\mathrm{rms}}_{cs},r^{\mathrm{rms}}_{\bar q_1 \bar q_2},r^{\mathrm{rms}}_{c\bar q_2},r^{\mathrm{rms}}_{s\bar q_1})$. This strongly suggests that $T_{cs,0(0)}^{\mathrm{Theo.}}(2350)$ is a $D\bar K$ molecular state. The $T_{cs,0(0)}^{\mathrm{Theo.}}(2350)$ exists below the $D\bar{K}$ threshold and can only decay weakly. The branching fractions for the Cabibbo-favored $c\to us\bar d$ decays of the charmed hadrons are usually on the order of 1\%~\cite{ParticleDataGroup:2022pth}. Therefore, the $T_{cs,0(0)}^{\mathrm{Theo.}}(2350)$ could be measured via the $B$ decays $B\to T_{cs}D$ with $T_{cs}\to K^-K^-\pi^+\pi^+$ in experiments~\cite{Yu:2017pmn, Chen:2020eyu}.

Below the $D^* \bar{K}^*$ threshold about 14 MeV, we detect a narrow resonance  $T_{cs,0(0)}^{\mathrm{Theo.}}(2906)$ with a width of $20$ MeV. Considering the theoretical errors of the AL1 model and the neglected $K^*$ widths, it stands as a good candidate for the experimental $T_{cs0}(2900)$. As listed in Table~\ref{tab:csmass_and_radii}, we can observe that the $r^{\mathrm{rms}}_{c\bar q_1}$ and $r^{\mathrm{rms}}_{s \bar q_2}$ of $T_{cs,0(0)}^{\mathrm{Theo.}}(2906)$ are smaller than the other radii in the four-body system and similar to the $D^*$ and $\bar{K}^*$ meson radii. This strongly indicates that $T_{cs,0(0)}^{\mathrm{Theo.}}(2906)$ is a molecular state of $D^*\bar{K}^*$. From Fig.~\ref{fig:Qsmass}(a1), one can see the pole is sandwiched between the two continuum lines of $D\bar{K}$ and $D^*\bar{K}^*$, which should be identified appearing in the RS-II of the $D\bar{K}$ channel and the RS-I of the $D^*\bar{K}^*$ channel. It could be regarded as a quasibound state of $D^*\bar{K}^*$ with its width decaying to $D\bar{K}$ as the imaginary part of the pole mass. The $T_{cs,0(0)}^{\mathrm{Theo.}}(2906)$ can be further measured in the $B\to T_{cs} D$ with $T_{cs}\to D^+K^-$. Additionally, the cross-verification can be performed through the $B\to T_{cs}\pi$ process~\cite{Chen:2020eyu}.

The $T^{\mathrm{Theo.}}_{cs,0(1)}(3431)$  is expected to be a diquark-antidiquark type resonance because of its relatively small $(r^{\mathrm{rms}}_{cs},r^{\mathrm{rms}}_{\bar q_1 \bar q_2})$. As for the other $T_{cs,I(J)}^{\mathrm{Theo.}}$ resonances in Table~\ref{tab:csmass_and_radii}, all of their radii are approximately around 1 fm, implying that they are the resonances with four (anti)quarks having similar roles.

The calculations from the chiral quark model~\cite{Yang:2021izl} obtain a deeply bound state with $I(J^P)=0(0^+)$ and do not yield states corresponding to the experimental observations. This discrepancy might be attributed to the excessive one-boson-exchange interactions in the model. In Ref.~\cite{Cheung:2020mql}, the lattice QCD simulations were performed with $m_\pi=239\mathrm{MeV}$ and $m_\pi=391\mathrm{MeV}$, and the hints of a $0(0^+)$ virtual-state pole below the $D \bar{K}$ threshold were identified. In Ref.~\cite{Ortega:2023azl}, the authors employed the chiral quark model and the resonating group method to investigate the $(cs\bar{q}\bar{q})$ states: they have identified poles close to the $D\bar{K}$ threshold and $D^*\bar{K}^*$ threshold, respectively, which are also dominated by the corresponding dimeson components. However, these poles are located in the different RSs compared with our results. The lower state is found in the RS-II of the $D\bar{K}$ channel and the RS-I of the $D^*\bar{K}^*$ channel, while the higher state is found in the RS-II of the $D\bar{K}$ channel and the RS-II of the $D^*\bar{K}^*$ channel. The results for the other quantum numbers are also inconsistent with our findings. Further experimental measurements are required to resolve the discrepancies arising from different models and methods.

\subsection{The $(bs\bar{q}\bar{q})$ sector}

For the $(bs\bar q\bar q)$ systems, we identify two shallow bound states, the $T_{bs,0(0)}^{\mathrm{Theo.}}(5781)$ with a binding energy of $3$ MeV and the $T_{bs,0(1)}^{\mathrm{Theo.}}(5840)$ with a binding energy of $1$ MeV. The radii presented in Table~\ref{tab:bsmass_and_radii} indicate that these two bound states are molecular states of $\bar{B}\bar{K}$ and $\bar{B}^*\bar{K}$, respectively. The fact that the $T_{bs,0(0)}^{\mathrm{Theo.}}(5781)$ and $T_{cs,0(0)}^{\mathrm{Theo.}}(2350)$ states share  almost the same binding energy also implies their molecular nature. In the molecular picture, the hadronic $\bar{B}\bar{K}$ interactions and the $D\bar{K}$ interaction are almost the same. Meanwhile, their reduced masses are mainly determined by the kaon mass because of the mass hierarchies of the heavy meson and kaon. Therefore, we get almost the same binding energy for these two systems. For the compact tetraquark bound states, replacing the $c$ quark with the $b$ quark would generally result in significantly deeper binding energies~\cite{Meng:2020knc,Meng:2023for,Meng:2021yjr,Meng:2023jqk,Ma:2023int}. The $T_{bs,0(0)}^{\mathrm{Theo.}}(5781)$ exists below the $\bar{B}\bar{K}$ threshold and can only decay weakly. The search for the $T_{bs,0(0)}^{\mathrm{Theo.}}(5781)$ could be performed in the processes of $T_{bs}\to J/\Psi K^- K^- \pi^+$ and $D^+K^-\pi^-$~\cite{Yu:2017pmn}. In addition to the weak decay processes, the $T_{bs,0(1)}^{\mathrm{Theo.}}(5840)$ could also be searched for by measuring the electromagnetic decay process like $T_{bs}\to \bar{B}^0 K^-\gamma$~\cite{Qin:2020zlg}.

Near the $\bar{B}^*\bar{K}^*$ threshold, we find a molecular resonance $T_{bs,0(0)}^{\mathrm{Theo.}}(6240)$ with a width of 18 MeV. This state serves as the partner of the $T_{cs,0(0)}^{\mathrm{Theo.}}(2906)$ when the $c$ quark is replaced by a $b$ quark. The $T_{bs,0(0)}^{\mathrm{Theo.}}(6240)$ and the $T_{cs,0(0)}^{\mathrm{Theo.}}(2906)$ share similar widths and have the same mass differences compared to their constituent mesons. This similarity further supports their molecular nature. The $T_{bs,0(0)}^{\mathrm{Theo.}}(6240)$ could be searched for via its strong decay process $T_{bs}\to \bar B^0 K^-$.

In Table~\ref{tab:bsmass_and_radii}, the other higher $T_{bs,I(J)}^{\mathrm{Theo.}}$ resonances are also presented. The radii indicate that they are resonances with four (anti)quarks playing similar roles. 

\section{Summary}\label{sec:summary}

In summary, we calculate the mass spectrum of the $Qs\bar q \bar q$ tetraquark states with $J^P=(0,1,2)^+$ using the AL1 quark potential model, known for its successful description of the conventional hadron spectrum. The potential parameters were determined through a best-fit procedure applied to a comprehensive set of mesons. This study does not introduce any additional free parameters. We employ the Gaussian expansion method to solve the four-body Schr\"odinger equation and the CSM to  distinguish the resonances from the scattering states.

We have identified several bound states and resonances in both the $(cs\bar q \bar q)$ and $(bs \bar q \bar q)$ sectors. Furthermore, we analyze the resulting spatial wave functions to determine the internal quark
clustering behavior of the identified states.

For the $(cs\bar q\bar q)$ systems, we obtain a molecular bound state $T_{cs,0(0)}^{\mathrm{Theo.}}(2350)$ close to the $D\bar{K}$ threshold, a molecular resonance $T_{cs,0(0)}^{\mathrm{Theo.}}(2906)$ close to the $D^*\bar{K}^*$ threshold, and several other higher $T_{cs,I(J)}^{\mathrm{Theo.}}$ resonances. Similarly, for the $(bs\bar q\bar q)$ systems, we have identified two molecular bound states $T_{bs,0(0)}^{\mathrm{Theo.}}(5781)$ close to the $\bar{B}\bar{K}$ threshold and $T_{bs,0(1)}^{\mathrm{Theo.}}(5840)$ close to the $\bar{B}^*\bar{K}$ threshold, along with a molecular resonance $T_{bs,0(0)}^{\mathrm{Theo.}}(6240)$ close to $\bar{B}^*K^*$ and several higher $T_{bs,I(J)}^{\mathrm{Theo.}}$ resonances. The $T_{cs,0(0)}^{\mathrm{Theo.}}(2906)$ near the $D^* \bar{K}^*$ threshold, with a width of $20$ MeV, stands as a good candidate for $T_{cs0}(2900)$. We also provide recommendations for experimental measurements of the predicted states.

While we reveal the molecular characteristics of some states through their energies and radii, these molecular states may disappear if we do not consider the diquark-antidiquark spatial configuration as shown in Fig.~\ref{fig:jac}. We conjecture that the matrix elements with the intermediate diquark-antidiquark states, e.g.,
$\langle \psi_\text{dimeson}|V|\psi_\text{diquark}\rangle\langle  \psi_\text{diquark} | V|\psi_\text{dimeson}\rangle $ are still very important to get the molecular states, where $\psi_\text{dimeson}$ and $\psi_\text{diquark}$ are the typical dimeson state and diquark-antidiquark state, respectively. The above conclusion is consistent with our previous works~\cite{Ma:2022vqf,Ma:2023int,Meng:2023jqk}, where the benchmark calculations with various quark potential models and few-body numerical methods were performed.

\section*{ACKNOWLEDGMENTS}
We thank Guang-Juan Wang, Yao Ma, Zi-Yang Lin, and Liang-Zhen Wen for the helpful discussions. This project was supported by the National
Natural Science Foundation of China (11975033 and 12070131001). This
project was also funded by the Deutsche Forschungsgemeinschaft (DFG,
German Research Foundation, Project ID 196253076-TRR 110). 

\bibliography{references}

\onecolumngrid
\clearpage
\twocolumngrid




\end{document}